\documentclass[11pt]{article}

\pdfoutput=1

\usepackage{amsmath,amssymb,amsfonts,amscd,mathrsfs}

\usepackage[utf8]{inputenc}

\usepackage[]{graphicx}
\usepackage{dsfont} 
\usepackage{verbatim}

\usepackage{xcolor}
\definecolor{darkblue}{rgb}{0.1,0.1,.7}
\usepackage[colorlinks, linkcolor=darkblue, citecolor=darkblue, urlcolor=darkblue, linktocpage]{hyperref} 
\usepackage[square, comma, compress,numbers]{natbib}
\usepackage{geometry}
\usepackage[margin=10pt,font=small,labelfont=bf]{caption}

\def\g{\gamma}

\def\bsub{\begin{subequations}}
\def\esub{\end{subequations}}
\def\zb{\bar{z}}

\def\lra{\leftrightarrow}

\def\dd{\delta}

\newcommand{\floor}[1]{\left \lfloor #1 \right \rfloor }

\newcommand{\ket}[1]{|#1\rangle}

\newcommand{\expec}[1]{\langle #1 \rangle}
\def\ldef{\mathrel{\mathop:}=}

\newcommand{\reef}[1]{(\ref{#1})}
\def\beq{\begin{equation}} 
\def\eeq{\end{equation}}

\def\mbb{\mathbb}
\def\mbf{\mathbf}
\def\mca{\mathcal}
\def\mfr{\mathfrak}
\def\mrm{\mathrm}

\def\pd{\partial}
\def\a{\alpha}
\def\b{\beta}

\def\la{\lambda}

\def\DD{\Delta}

\def\Oo{\mathcal{O}}
\def\sO{\mathrm{O}}

\def\l{\ell}

\newcommand{\limu}[1]{\mathrel{\mathop{\sim}\limits_{\scriptstyle{#1}}}}

\numberwithin{equation}{section}

\vspace{-2cm}

\begin{document}

\vspace*{-.6in} \thispagestyle{empty}
\begin{flushright}
YITP-SB-16-11
\end{flushright}
\vspace{1cm} {\Large
\begin{center}
  {\bf Dimensional Reduction for Conformal Blocks}
\end{center}}
\vspace{1cm}
\begin{center}
{\bf Matthijs Hogervorst}\\[1cm] 
{
C.N.\@ Yang Institute for Theoretical Physics, Stony Brook University, USA
}
\\
\end{center}
\vspace{14mm}

\begin{abstract}
  We consider the dimensional reduction of a CFT, breaking multiplets of the $d$-dimensional conformal group $SO(d+1,1)$ up into multiplets of $SO(d,1)$. This leads to an expansion of $d$-dimensional conformal blocks in terms of blocks in $d\!-\!1$ dimensions. In particular, we obtain a formula for 3$d$ conformal blocks as an infinite sum over ${}_2F_1$ hypergeometric functions with closed-form coefficients.
\end{abstract}
\vspace{12mm}
\hspace{7mm}April 2016

\newpage

{
\setlength{\parskip}{0.05in}
\tableofcontents
}

\setlength{\parskip}{0.05in}

\section{Introduction}

Conformal Field Theories (CFTs) in $d \geq 2$ spacetime dimensions are being intensively studied via the bootstrap program, revived in Ref.~\cite{Rattazzi:2008pe}. The conformal bootstrap has led to a deeper analytic understanding of (super)conformal field theories~\cite{Fitzpatrick:2012yx,Komargodski:2012ek,Beem:2013sza,Beem:2014zpa,Beem:2015aoa,Hartman:2015lfa,Hofman:2016awc} but also to precise numerical predictions for critical exponents, see e.g.~\cite{El-Showk:2014dwa,Kos:2016ysd}. A pedagogical treatment of the subject is given in~\cite{Rychkov:2016iqz,Simmons-Duffin:2016gjk}.

A crucial role in this program is played by \emph{conformal blocks}, special functions that are determined completely by conformal symmetry.
They were first described in the 1970s~\cite{Polyakov:1974gs,Ferrara:1973vz,Ferrara:1974nf,Ferrara:1974ny} but have received much attention in recent years after breakthrough results by Dolan and Osborn~\cite{DO1,DO2,DO3}. Currently, simple expressions for conformal blocks are only known in even spacetime dimension $d$. Recent work has led to systematic methods for computing these blocks in any $d$~\cite{Hogervorst:2013sma,Hogervorst:2013kva,Kos:2013tga,Kos:2014bka}. 
Moreover, much is now known about conformal blocks that appear in four-point functions with spinning operators~\cite{Costa:2011dw,SimmonsDuffin:2012uy,Echeverri:2015rwa,Iliesiu:2015qra,Rejon-Barrera:2015bpa,Iliesiu:2015akf,Penedones:2015aga,Echeverri:2016dun,Costa:2016hju,Costa:2016xah} and about superconformal blocks, see e.g.~\cite{Fitzpatrick:2014oza}.

In this note we develop a new representation of conformal blocks in $d$ dimensions. This representation arises from the ``dimensional reduction'' of a CFT, i.e.\@ the restriction of the conformal group $SO(d+1,1)$ to a subgroup $SO(d,1)$ that preserves a hyperplane of codimension one. Although this is similar in spirit to a Kaluza-Klein reduction, we are not actually truncating the theory: rather, we simply organize all states in the Hilbert space of the CFT in representations of $SO(d,1)$ instead of the full conformal group. In particular, a $d$-dimensional conformal block will decompose into infinitely many $(d\!-\!1)$-dimensional conformal blocks with computable coefficients. As a corollary, this strategy provides an explicit formula for 3$d$ and 5$d$ conformal blocks in terms of ${}_2F_1$ hypergeometric functions.

This paper is organized as follows. Section~\ref{sec:main} reviews basic facts about conformal blocks and develops the promised dimensional reduction. In section~\ref{sec:lightcone}, we compare our expansion in $d-1$ dimensional blocks to an expansion in 2$d$ blocks. Finally section~\ref{sec:disc} discusses several directions for future work. Appendix~\ref{sec:free} is a consistency check of the formalism developed in this note, applying it to the four-point function of the free scalar field.

\section{Dimensional reduction}\label{sec:main}
Let's start by recalling the definition of conformal blocks. For concreteness, consider a scalar operator $\phi$ of scaling dimension $\DD_\phi$ in a unitary $d$-dimensional CFT. Conformal invariance requires that its four-point function is of the following form: 
\beq
\label{eq:4pt}
 \expec{\phi(x_1)\phi(x_2)\phi(x_3)\phi(x_4)} = \frac{\mca{G}_\phi(u,v)}{|x_1 - x_2|^{2\DD_\phi}|x_3 - x_4|^{2\DD_\phi}}
 \eeq
 where the function $\mca{G}_\phi(u,v)$ depends only on two conformally invariant cross ratios
\beq
\hspace{10mm} u = \frac{x_{12}^2 x_{34}^2}{x_{13}^2 x_{24}^2}\,, \qquad v = \frac{x_{14}^2 x_{23}^2}{x_{13}^2 x_{24}^2}\,, \hspace{20mm} x_{ij} \ldef x_i - x_j\,.
\eeq
The four-point function~\reef{eq:4pt} can be computed using the operator product expansion (OPE):
\beq
\phi(x)\phi(0) = \frac{1}{|x|^{2\DD_\phi}} + \sum_{\Oo = [\DD,\l]} \frac{\la_\Oo}{|x|^{2\DD_\phi - \DD}} \, C_\DD^{(\l)}(x,\pd)^{\mu_1 \dotsm \mu_\l} \Oo_{\mu_1 \dotsm \mu_\l}(0)\,.
\eeq
Here the sum runs over all primary operators $\Oo_{\mu_1 \dots \mu_\l}(x)$ of even spin $\l$ in the theory; with $\DD$ we denote their scaling dimension, and the OPE coefficient $\la_\Oo$ is the constant of proportionality appearing in the three-point function $\expec{\phi \phi \Oo}$. The differential operator $C_\DD^{(\l)}(x,\pd)^{\mu_1 \dotsm \mu_\l}$ depends only on the quantum numbers $\DD$ and $\l$. In passing, we note that unitary puts a lower bound on the possible values that $\DD$ can have: 
\beq
\label{eq:unitbd}
\DD \; \geq \; \begin{cases} (d-2)/2 \quad & \l = 0 \\ \l + d - 2 & \l \geq 1 \end{cases}\;.
\eeq
By applying the OPE twice to the four-point function~\reef{eq:4pt}, one can show that $\mca{G}_\phi(u,v)$ can be written as follows:
\beq
\label{eq:CBdec}
\mca{G}_\phi(u,v) = 1 + \sum_{\Oo = [\DD,\l]} (\la_\Oo)^2 \; G_{\DD}^{(\l)}(u,v;d)\,.
\eeq
The functions $G_\DD^{(\l)}(u,v;d)$ are conformal blocks, hence Eq.~\reef{eq:CBdec} is known as a conformal block (CB) decomposition. As the notation indicates, the blocks only depend on the quantum numbers $\DD$ and $\l$  and the spacetime dimension $d$. In practice, they can be computed by solving a second-order PDE~\cite{DO2} while imposing the following asymptotic behaviour:
\beq
\label{eq:asymp}
G_\DD^{(\l)}(u,v;d) \; \limu{u \to 0,\, v \to 1} c_\l^{(d)} \;  u^{\DD/2} \, \widehat{C}_\l^{(\nu)}\!\left(\frac{1-v}{2\sqrt{u}}\right),
\eeq
where $\widehat{C}_j^{(\nu)}$ is a rescaled Gegenbauer polynomial with parameter $\nu \ldef (d-2)/2$:
\beq
\widehat{C}_j^{(\nu)}(\xi) \ldef \frac{j!}{(2\nu)_j} \mrm{Geg}_j^{(\nu)}(\xi), \qquad (x)_n \ldef \Gamma(x+n)/\Gamma(x)\,.
\eeq
By construction, these functions obey $\widehat{C}_j^{(\nu)}(1) = 1$ and have a finite limit as $d \to 2$, contrary to the normal Gegenbauer polynomials. A natural choice for the normalization coefficients $c_\l^{(d)}$ is~\cite{Costa:2011dw} 
\beq
\label{eq:omegaDef}
c_\l^{(d)}  = \frac{(-1)^\l (2\nu)_\l}{2^\l (\nu)_\l}
\eeq
although we will leave $c_\l^{(d)}$ arbitrary in the rest of this paper.

In even spacetime dimensions, simple expressions for the conformal blocks exist~\cite{DO1,DO2,DO3}. These are easiest to state in the Dolan-Osborn coordinates $z,\zb$, defined through $u = z\zb$, ${v = (1-z)(1-\zb)}$. On the Euclidean section, $z$ is a complex coordinate and $\zb = z^*$ its conjugate. 
After defining 
\beq
k_{2\beta}(x) \ldef x^{\beta} \, {}_2F_1\!\left( \b,\b;2\b;x\right)
\eeq
the 2$d$ and 4$d$ conformal blocks are:
\bsub
\label{eq:even}
\begin{align}
\label{eq:2d}
G_\DD^{(\l)}(z,\zb;2) &= \frac{c_\l^{(2)}}{2} \, \big[ k_{\DD+\l}(z) k_{\DD - \l}(\zb) \; + \; (z \lra \zb) \big] \\
\label{eq:4d}
G_\DD^{(\l)}(z,\zb;4) &= \frac{c_\l^{(4)}}{\l+1} \frac{z\zb}{z-\zb} \big[ k_{\DD+\l}(z) k_{\DD - \l-2}(\zb) \; - \; (z \lra \zb) \big]\,.
\end{align}
\esub
No similar formulas in odd $d$ are known, although some simplifications occur when specializing to the ``diagonal'' line $z = \zb$~\cite{Hogervorst:2013kva,Rychkov:2015lca}.

The conformal block $G_\DD^{(\l)}$ has a representation-theoretical meaning: it is the contribution of a conformal multiplet of dimension $\DD$ and spin $\l$ to the four-point function~\reef{eq:4pt}, containing a primary operator $\Oo_{\mu_1 \dotsm \mu_\l}(x)$ and all of its derivatives. Such a multiplet can be described in a concrete fashion through the state-operator correspondence. The multiplet of $\Oo$ is built on top of the primary state $\ket{\Oo}_{\mu_1 \dotsm \mu_\l} \ldef \lim_{x \to 0} \Oo_{\mu_1 \dots \mu_\l}(x) \ket{0}$, where $\ket{0}$ is the CFT vacuum. All other states in the multiplet are obtained by acting on $\ket{\Oo}$ with $P_\mu$, the generator of translations of the conformal algebra. A complete basis\footnote{We are ignoring descendants that transform in mixed or antisymmetric representations of the Lorentz group, since such descendants do not contribute to a scalar four-point function.} of these descendants is spanned by the following states:
\beq
\label{eq:descstate}
(P^2)^k P_{\mu_1} \dotsm P_{\mu_m} P^{\nu_1} \dotsm P^{\nu_p} \ket{\Oo}_{\nu_1 \dotsm \nu_p \, \mu_{m+1} \dotsm \mu_{m+r}}\,, \qquad r = \l - p\,, \quad 0 \leq p \leq \l\,.
\eeq
It is understood that the $\mu$ indices must be symmetrized and made traceless. The state shown in~\reef{eq:descstate} then has scaling dimension $\DD + 2k + m + p$ and spin $\l +m - p$.  It follows that a descendant of level $n$ --- that is to say, with dimension $\DD + n$ --- can have the following spins:
\beq
\label{eq:spinRange}
j = \l + n, \, \l + n -2,\, \ldots,\, \max( \l-n,\, \l-n \text{ mod }2 ).
\eeq
For a suitable choice of coordinates, there is one-to-one correspondence between a descendant of level $n$ and spin $j$ and a term in the conformal block $G_\DD^{(\l)}$. To make this concrete, we pass to the following coordinates:
\beq
\label{eq:sDef}
s \ldef |z| = \sqrt{z \zb}\,, \quad \xi \ldef \cos(\arg z) = \frac{z + \zb}{2 \sqrt{z \zb}}\,.
\eeq
In the $(s,\xi)$ coordinates, the contribution of a level-$n$ spin-$j$ descendant to the conformal block can be shown~\cite{Hogervorst:2013sma} to be proportional to $\mbf{P}_{\DD+n,j}^{(d)}(s,\xi)$, where
\beq
\mbf{P}_{E,j}^{(d)}(s,\xi) \ldef s^E \, \widehat{C}_j^{(\nu)}(\xi)\,.
\eeq
This is consistent with the fact that Gegenbauer polynomials are $d$-dimensional spherical harmonics. Consequently, conformal blocks admit an expansion of the form
\beq
\label{eq:serRep}
G_{\DD}^{(\l)}(s,\xi;d) = \sum_{n=0}^{\infty} \sum_{j}  a_{n,j}^{(d)}(\DD,\l) \, \mbf{P}_{\DD+n,j}^{(d)}(s,\xi) 
\eeq
with $j$ again restricted to the range~\reef{eq:spinRange}. The coefficients $a_{n,j}^{(d)}$ are fixed by conformal invariance, and are known in closed form as ${}_4F_3$ hypergeometrics evaluated at unity~\cite{DO2}.

As advertised, we will break the conformal group down to a subgroup of conformal transformations in $d-1$ dimensions, and we want to analyze the consequences of this dimensional reduction for conformal blocks. Let us first consider a toy example of what will happen, namely the restriction of the rotation group $SO(d)$ to $SO(d\!-\!1)$. If we think of $SO(d)$ as the isometry group of the sphere $S^{d-1}$, this means that we take the subgroup of rotations that leave the equator invariant. Under this restriction, the spin-$\l$ representation of $SO(d)$, denoted as $[\l]_d$, breaks up into $SO(d\!-\!1)$ irreps as follows:
\beq
\label{eq:branching}
[\l]_d = [0]_{d-1}  +  [1]_{d-1}  + \ldots +  [\l]_{d-1}\,.
\eeq
The branching rule~\reef{eq:branching} can be understood by realizing $[\l]_d$ as a traceless symmetric tensor of rank $\l$. For instance, the first $d-1$ components of a vector $v_\mu \in \mbb{R}^d$ form a vector representation of ${SO(d\!-\!1)}$, whereas the last component $v_d$ transforms as a $SO(d\!-\!1)$ scalar.

Since spherical harmonics form a representation of $SO(d)$, the branching rule~\reef{eq:branching}  applies in particular to the (rescaled) Gegenbauer polynomials. Concretely, the spin-$\l$ Gegenbauer polynomial $\widehat{C}_\l^{(\nu)}$ can be written in the following form:
\beq
\label{eq:gegDec}
\widehat{C}_\l^{(\nu)}(\xi) = \sum_{j=0}^\l Z_{\l}^{j} \, \widehat{C}_j^{(\nu-1/2)}(\xi)
\eeq
since the $\widehat{C}_j^{(\nu-1/2)}$ are Gegenbauer polynomials in $d-1$ dimensions. 
As a matter of fact, only spins $j = \l,\,\l-2,\ldots, \, \l \text{ mod } 2$ appear in the RHS of Eq.~\reef{eq:gegDec}, owing to the selection rule
\beq
\widehat{C}_j^{(\nu)}(-\xi) = (-1)^j \, \widehat{C}_j^{(\nu)}(\xi)\,.
\eeq
The coefficients $Z_\l^j$ in Eq.~\reef{eq:gegDec} can be computed using explicit expressions for the Gegenbauer polynomials~\cite{Bateman1} together with their orthogonality. This yields
\beq
\label{eq:gegId}
 Z_{\l}^{j} = \frac{(1/2)_p \, \l! }{ p!\,j! }  \frac{(\nu)_{j+p}(2\nu-1)_j}{(\nu-1/2)_{j+p+1}(2\nu)_\l}(j+\nu-1/2)\,, \qquad p \equiv (\l-j)/2\,.
\eeq
It will prove useful later in this work to have a bound on the coefficients $Z_\l^j$. It is easy to see that all $Z_\l^j$ are  positive, provided that $d \geq 2$.\footnote{This is a special case of the fact that Gegenbauer polynomials in $D$ dimensions can be written as a sum over Gegenbauer polynomials in $d < D$ dimensions with positive coefficients, see~\cite{askey} and references therein. We thank A.\@ Zhiboedov for pointing this out.} Moreover, the normalization condition $\widehat{C}_\l^{(\nu)}(1) = 1$ implies that for fixed $\l$ we have
\beq
\sum_j Z_\l^j = 1\,.
\eeq
We conclude that $0 \leq Z_\l^j \leq 1$ for all $\l,j$. 

Having considered the restriction $SO(d) \to SO(d\!-\!1)$, we now turn our attention to the conformal group $SO(d+1,1)$. We will restrict the full group to a subgroup $SO(d,1)$ that preserves the hyperplane $x_d = 0$. Doing so, a primary $d$-dimensional representation breaks up into infinitely many primary $(d\!-\!1)$-dimensional representations. The argument is the following. Recall that a state is a primary of $SO(d+1,1)$ if and only if it is annihilated by all $d$ generators of special conformal transformations, which we denote here by $K_\mu$. Therefore any state that is annihilated by $K_{1},\ldots,K_{d-1}$ but not by $K_d$ is a descendant of $SO(d+1,1)$ but a primary of $SO(d,1)$. Among all descendants shown in Eq.~\reef{eq:descstate}, the following states fit that description:
\beq
\label{eq:branch2}
\ket{\Oo;j,m}_{\a_1 \dotsm \a_j} = (P_d)^m \ket{\Oo}_{\a_1 \dotsm \a_j \,  d \dotsm d}\,, \qquad 0 \leq j \leq \l\,, \; m = 0,1,2,\ldots.
\eeq
The state $\ket{\Oo;j,m}$ has $SO(d-1)$ spin $j$ and scaling dimension $\DD + m$. We arrive at the following branching rule: any $SO(d+1,1)$ multiplet of dimension $\DD$ and spin $\l$ splits up into infinitely many $SO(d,1)$ multiplets of spin $0 \leq j \leq \l$ and dimension $\DD + m$ with $m \geq 0$.\footnote{
Such branching rules can also be derived or checked by decomposing the characters of $SO(d+1,1)$ into $SO(d,1)$ characters, see e.g.~\cite{Dolan:2005wy}. This approach may be useful when dealing with more complicated representations.}

Consequently, a conformal block $G_{\DD}^{(\l)}(u,v;d)$ can be written as an infinite sum over the conformal blocks $G_{\DD+m}^{(j)}(u,v;d-1)$ with $0 \leq j \leq \l$ and $m \geq 0$. There are however some selection rules that apply, as was the case for the Gegenbauer polynomials. We will derive these in the $\rho$ kinematics of~\cite{Pappadopulo:2012jk}, passing to the $(r,\eta)$ coordinates defined as
\beq
u = \left(\frac{4r}{1 + 2 r \eta + r^2}\right)^2\,,
\quad
v = \left(\frac{1 - 2r \eta + r^2}{1 + 2 r \eta + r^2} \right)^2\,.
\eeq
In the $(r,\eta)$ coordinates, conformal blocks have an expansion where only descendants of \emph{even} level appear~\cite{Hogervorst:2013sma}:
\beq
\label{eq:rhoSeries}
G_{\DD}^{(\l)}(r,\eta;d) =  \sum_{n=0}^\infty \sum_j \, b_{n,j}^{(d)}(\DD,\l) \, \mbf{P}^{(d)}_{\DD + 2n,j}(r,\eta)
\eeq
with $j$ restricted to the range~\reef{eq:spinRange}. 
The $SO(d+1,1) \to SO(d,1)$ branching rule described above must apply to any coordinate set, in particular to the $(r,\eta)$ coordinates. By consistency with Eq.~\reef{eq:rhoSeries}, it follows that only $SO(d,1)$ primaries of even level can appear in the decomposition of $G_\DD^{(\l)}(u,v;d)$. Likewise, only spins $j = \l, \, \l-2,\ldots,\ \l \text{ mod } 2$ can contribute. In conclusion, there exists a decomposition of $G_\DD^{(\l)}$ of the following form: 
\beq
\label{eq:genRep}
G_{\DD}^{(\l)}(u,v;d) = \sum_{n=0}^{\infty} \sum_{j} \mca{A}_{n,j}(\DD,\l) \, G_{\DD+2n}^{(j)}(u,v;d-1)
\eeq
with the sum running over
\beq
j = \l\,, \, \l -2\,, \ldots, \, \l \text{ mod } 2\,.
\eeq
The coefficients $\mca{A}_{n,j}(\DD,\l)$ are fixed by conformal invariance. In the following section, we will explain one method to compute them. 

\subsection{Recursion relation for coefficients}

In this section, we will compute the coefficients $\mca{A}_{n,j}(\DD,\l)$ appearing in Eq.~\reef{eq:genRep}.  Our discussion will rely heavily on the representation~\reef{eq:serRep} of conformal blocks in the $(s,\xi)$ coordinates. In particular, we will use the fact that the coefficients $a_{n,j}^{(d)}(\DD,\l)$ obey a three-term recursion relation:
\begin{multline}
\label{eq:aRR}
\left[\mrm{C}^{(d)}(\DD+n,j) - \mrm{C}^{(d)}(\DD,\l)\right]\!a_{n,j}^{(d)}(\DD,\l) \\
=  \; \b^{(d)}_{j-1}(\DD+n-1) \, a_{n-1,j-1}^{(d)}(\DD,\l) \, + \,  \g^{(d)}_{j+1}(\DD+n-1) \, a_{n-1,j+1}^{(d)}(\DD,\l)\,.
\end{multline}
Here $\mrm{C}^{(d)}(\DD,\l) \ldef \DD(\DD-d) + \l(\l+d-2)$ is the eigenvalue of the quadratic conformal Casimir and
\beq
\b^{(d)}_{j}(x) \ldef \frac{(x+j)^2 (j+2\nu)}{2(j+\nu)}\,, \quad
\g^{(d)}_{j}(x) \ldef \frac{(x-j-2\nu)^2j}{2(j+\nu)}\,.
\eeq
We can use this recurrence to compute the coefficients $a_{n,j}^{(d)}(\DD,\l)$ to arbitrary order in $n$, starting from the initial condition
\beq
\label{eq:initC}
a_{0,j}^{(d)}(\DD,\l) = c_\l^{(d)} \, \dd_{j,\l}
\eeq
that is imposed by Eq.~\reef{eq:asymp}. A comprehensive discussion of this recursion relation is given in~\cite{Hogervorst:2013sma}. 

We will compute the $\mca{A}_{n,j}(\DD,\l)$ coefficients by formulating a second recursion relation. As a starting point, remark that the conformal block $G_{\DD}^{(\l)}(s,\xi;d)$ admits an expansion in terms of the functions $\mbf{P}_{\DD+n,j}^{(d-1)}(s,\xi)$. 
This expansion takes the following form:
\beq
\label{eq:newRep}
G_{\DD}^{(\l)}(s,\xi;d) = \sum_{n = 0}^\infty \sum_{j} Y^{(\l)}_{n,j}\, \mbf{P}^{(d-1)}_{\DD+n,j}(s,\xi)\,, \qquad j \in \{ \l + n,\, \l + n-2, \ldots, \, \l \text{ mod } 2 \}
\eeq
for some coefficients $Y^{(\l)}_{n,j}$ that we will determine. On the one hand, Eq.~\reef{eq:newRep} can be obtained by applying the Gegenbauer identity~\reef{eq:gegId} to the $(s,\xi)$ representation of Eq.~\reef{eq:serRep}. We can therefore express the coefficients $Y^{(\l)}_{n,j}$ as follows:
\beq
Y_{n,j}^{(\l)} = \sum_{p=0}^{p_*} Z_{j+2p}^{j} \, a_{n,j+2p}^{(d)}(\DD,\l), \qquad p_* = (\l + n - j)/2\,.
\eeq
On the other hand, we can first apply the dimensional reduction formula~\reef{eq:genRep} to the conformal block $G_\DD^{(\l)}(u,v;d)$. Next, we expand every $(d\!-\!1)$-dimensional block on the RHS in terms of the $(s,\xi)$ representation. Doing so leads to a different expression for the $Y_{n,j}^{(\l)}$, namely
\beq
Y_{n,j}^{(\l)} = \sum_{m=0}^{\floor{n/2}} \sum_{k=0}^\l  \mca{A}_{m,k}(\DD,\l) \, a_{n-2m,j}^{(d-1)}(\DD+2m,k) \,.
\eeq
The sum over $k$ is also restricted to $j + 2m - n \leq k \leq j + n -2m$ and $ k \equiv \l \text{ mod } 2$. 

Now fix $j \in \{ \l, \, \l-2, \ldots,\, \l \text{ mod } 2\}$ and $n \geq 0$. Requiring that the two expressions for $Y_{2n,j}^{(\l)}$ agree, we obtain the following identity:
\beq
\label{eq:Arecrel}
c_j^{(d-1)}  \mca{A}_{n,j}(\DD,\l) = \sum_{q=0}^{q_*}    \; Z_{j+2q}^{j} \, a_{2n,j+2q}^{(d)}(\DD,\l)  -  \sum_{m=0}^{n-1} \sum_{k} \mca{A}_{m,k}(\DD,\l) \, a_{2n-2m,j}^{(d-1)}(\DD+2m,k)
\eeq
where $q_* = (\l + 2n - j)/2$ and $k$ is restricted to
\beq
\max(0,j + 2m - 2n) \leq k \leq \min(\l,j + 2n -2m), \quad k \equiv \l \text{ mod } 2\,.
\eeq
Notice that the RHS of~\reef{eq:Arecrel} only involves coefficients $\mca{A}_{m,k}(\DD,\l)$ with $m < n$. Moreover, the coefficients $a_{2n,j+2q}^{(d)}(\DD,\l)$ and $a_{2n-2m,j}^{(d-1)}(\DD+2m,k)$ can be computed by means of the recursion relation~\reef{eq:aRR}. We can therefore use Eq.~\reef{eq:Arecrel} to compute the coefficients $\mca{A}_{n,j}(\DD,\l)$ recursively, up to arbitrary $n$, starting from $n=0$. To be precise, Eq,~\reef{eq:Arecrel} must be understood as a set of $\floor{\l/2}$ coupled recursion relations, one for every allowed value of $j$. Finally, we remark that the above recursion relation is inhomogeneous, which means that the ``initial condition'' $\mca{A}_{0,j}(\DD,\l)$ is not arbitrary. Concretely, setting $n=0$ in Eq.~\reef{eq:Arecrel} yields
\beq
\mca{A}_{0,j}(\DD,\l) = Z_\l^j \frac{c_\l^{(d)}}{c_j^{(d-1)}}
\eeq
which is consistent with the asymptotics imposed by Eq.~\reef{eq:asymp}.

Although the recursion relation~\reef{eq:Arecrel} looks complicated, its solution can be written down in closed form:
\begin{multline}
\label{eq:master}
  \mca{A}_{n,j}(\DD,\l) = Z_\l^j \,\frac{c_\l^{(d)}}{c_j^{(d-1)}} \frac{ \big((\DD+j)/2 \big)_n\big((\tau+\l-j+1)/2 \big)_n }{\big((\DD+j-1)/2 \big)_n\big((\tau+\l-j)/2 \big)_n} \\
\times \; \frac{(1/2)_n}{16^n\, n!} \frac{(\DD-1)_{2n} \big( (\DD+\l)/2 \big)_n \left(\tau/2\right)_n }{(\DD - \nu)_n (\DD - \nu -1/2 +n)_n \big( (\DD + \l+1)/2 \big)_n \big( (\tau+1)/2 \big)_n}\,, 
\end{multline}
writing  $\tau \ldef \DD - (\l + d-2)$ for the conformal twist. While we don't have a rigorous proof of this formula, we have checked that it satisfies~\reef{eq:Arecrel} for $\l,n \leq 20$ and we conjecture that it holds in general. Clearly Eq.~\reef{eq:master} can be checked in other ways, e.g.\@ using expansions of conformal blocks in the $z$ or $\rho$ coordinates and in the diagonal limit.  

In passing, we notice that for the scalar ($\l=0$) block, only terms with $j=0$ are allowed in~\reef{eq:genRep}, and the formula for the coefficients simplifies:
\beq
\label{eq:scalar}
\mca{A}_{n,0}(\DD,0) =   \frac{c_0^{(d)}}{c_0^{(d-1)}} \frac{(1/2)_n}{4^n\, n!} \frac{(\DD/2)_{n}^3 }{(\DD - \nu)_n (\DD - \nu -1/2 +n)_n \big( (\DD +1)/2 \big)_n}\,.
\eeq
A similar simplication occurs for $\l = 1$.

\subsection{Convergence}

Equations~\reef{eq:genRep} and~\reef{eq:master} are the main result of this note. At this stage, we want to point out two important properties of the coefficients $\mca{A}_{n,j}$. For convenience, we will set $c_\l^{(d)} \equiv 1$ in what follows. First, we note that all $\mca{A}_{n,j}(\DD,\l)$ are positive, provided that $\DD$ satisfies the unitarity bound~\reef{eq:unitbd} and $d \geq 2$. Second, we remark that $\mca{A}_{n,j}$ decays exponentially fast with $n$. 
To prove this, let's consider the coefficient $\mca{A}_{n,j}(\DD,\l)$ as a function of $\DD$, keeping $\l,j$ and $n$ fixed. We notice that $\mca{A}_{n,j}(\DD,\l)$ is a rational function of $\DD$ of the form $p(\DD)/q(\DD)$, where $p$ and $q$ are polynomials of equal degree. Furthermore $p$ and $q$ completely factorize over the reals, with all zeroes at values of $\DD$ at or below the unitarity bound. 
This means that \emph{above} the unitarity bound, $\mca{A}_{n,j}(\DD,\l)$ is a slowly varying function of $\DD$, and it is well approximated by its value in the limit $\DD \to \infty$:
\beq
\label{eq:rate}
\mca{A}_{n,j}(\DD,\l) \, \limu{\DD \to \infty} \,   Z_\l^j \, \frac{(1/2)_n}{16^n \, n!}\left[ 1 + \sO\!\left(\frac{1}{\DD}\right) \right].
\eeq
As promised, the coefficient $\mca{A}_{n,j}(\DD,\l)$ decreases exponentially with $n$, as $\sim n^{-1/2} \, 16^{-n}$. Remarkably, this exponential behaviour holds not only asymptotically, but already starts at $n=1$.

\begin{figure}[htbp]
  \begin{center}
   \hspace{-12mm}\centerline{
\includegraphics[scale=0.65]{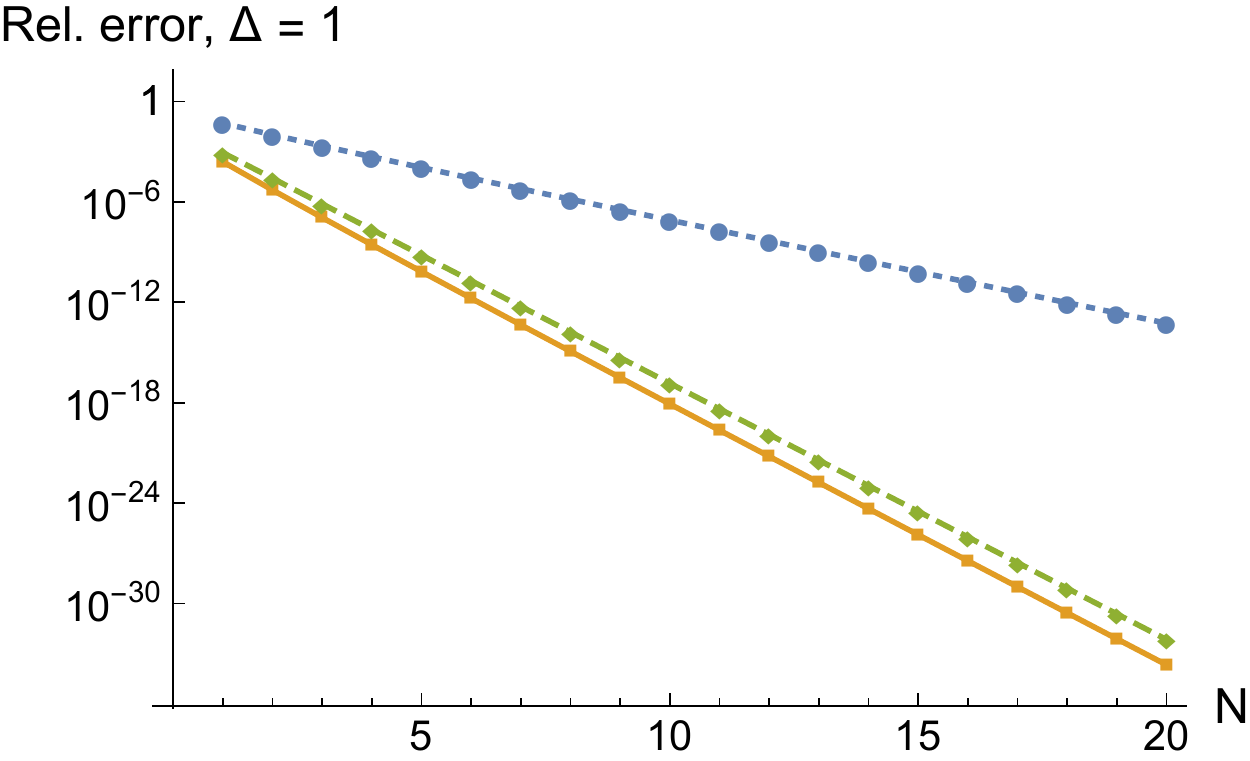} \;
\includegraphics[scale=0.65]{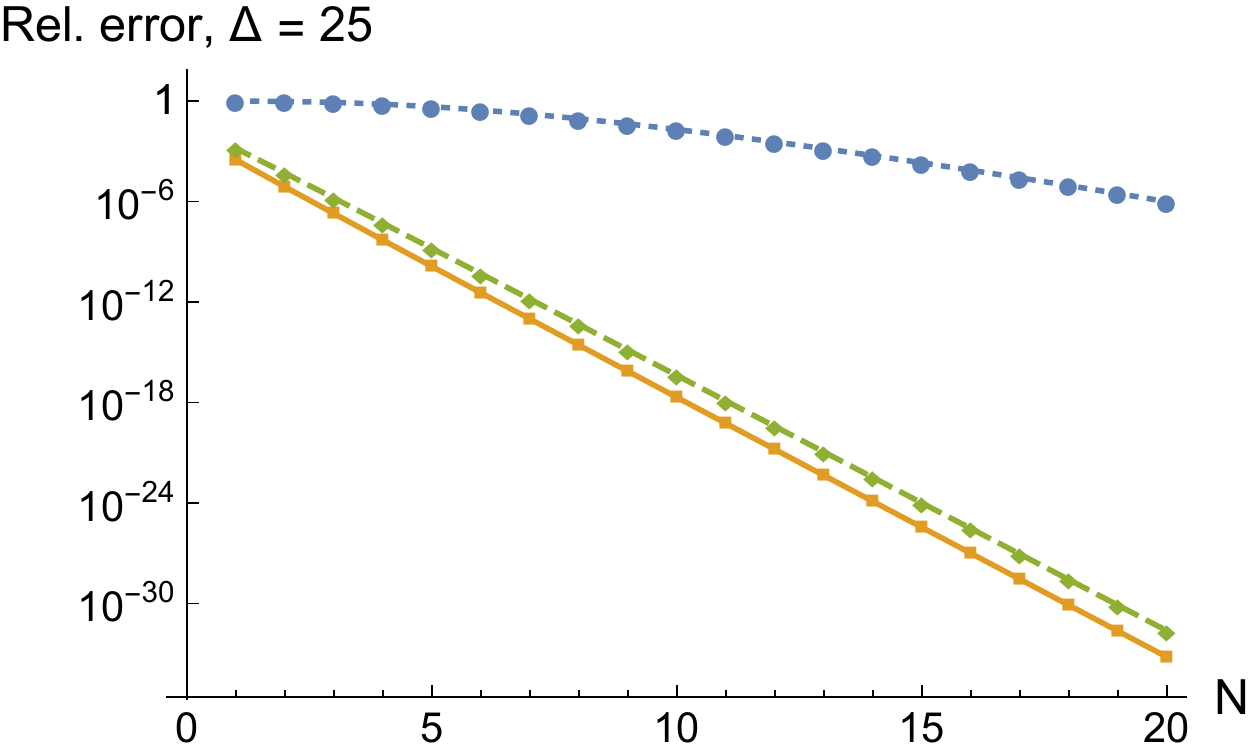}
    }
\caption{Comparison of different conformal block expansions. Horizontal axis: the order of truncation $N$, vertical axis: relative error in the numerical value of the conformal block --- notice the logarithmic scale. Solid orange: dimensional reduction with $n \leq N$ terms; dashed green: $\rho$-series with $n \leq N$ terms; dotted blue: $z$-series with $n \leq 2N$ terms. The points are joined by lines to guide the eye. The left plot shows the 3$d$ scalar block at the point $u=v=1/4$ with $\DD=1$, the right plot corresponds to $\DD = 25$.}
\label{fig:conv}
\end{center}
\end{figure}

So far, we have encountered three different expansions for $d$-dimensional conformal blocks: the ``$z$-series'' from Eq.~\reef{eq:serRep}, the ``$\rho$-series'' from~\reef{eq:rhoSeries} and the expansion in terms of lower-dimensional blocks~\reef{eq:genRep}. In Fig.~\ref{fig:conv} we compare their convergence rates numerically, by truncating these expansions at finite order $N$ and evaluating them at the crossing symmetric point ${u=v=1/4}$. The results corroborate that the truncation error of~\reef{eq:genRep} decreases exponentially with $N$.

For completeness, we can verify that the exponential decay with $n$ also holds for $\DD$ close to the unitarity bound. For spinning operators $(\l \geq 1)$, the limit $\tau \to 0$ is continuous, meaning that there are no important corrections to~\reef{eq:rate}, and the exponential decay at large $n$ persists. This is confirmed by an explicit expression for $\mca{A}_{n,j}$ at $\tau = 0$ shown in Appendix~\ref{sec:free}. As is well known, the scalar ($\l = 0$) block diverges at the unitarity bound $\DD = \nu$, where a level-two descendant becomes null. Using a conformal representation theory argument~\cite{zamolodchikov1984,Zamolodchikov:1987,Kos:2013tga,Penedones:2015aga}, we have
\beq
G_{\DD}^{(0)}(u,v;d) \; \limu{\DD \to \nu} \; \frac{1}{\DD - \nu} \frac{\nu^3}{16 (\nu+1)} \, G_{d/2+1}^{(0)}(u,v;d) \, + \, \ldots
\eeq
omitting terms that are regular as $\DD \to \nu$. Hence near the unitarity bound, $G_\DD^{(0)}$ is dominated by a conformal block with $\DD = d/2+1$, which itself is well above the unitarity bound. Therefore the estimate~\reef{eq:rate} applies, and we are done. The same conclusion can be reached by expanding Eq.~\reef{eq:scalar} around $\DD = \nu$. 

\subsection{Comparison to 2$d$ expansion}\label{sec:lightcone}

The dimensional reduction discussed in this paper has a counterpart on the lightcone, i.e.\@ the Minkowski section of a CFT, where $z$ and $\zb$ are independent, real variables. Lightcone kinematics turn out to be particularly simple:
in the limit $z \to 0$ at fixed $\zb$ the conformal blocks become effectively two-dimensional, up to an unimportant prefactor:
\beq
\label{eq:leadingLC}
G_\DD^{(\l)}(z,\zb;d) \; \limu{z \to 0} \; c_\l^{(d)} \frac{(\nu)_\l}{(2\nu)_\l} \,z^{(\DD-\l)/2} \, k_{\DD+\l}(\zb) \,.
\eeq
The study of CFT crossing equations in this limit has led to many analytic bootstrap results, initiated in~\cite{Fitzpatrick:2012yx,Komargodski:2012ek} with follow-up work in Refs.~\cite{Alday:2013cwa,Costa:2013zra,Fitzpatrick:2014vua,Vos:2014pqa,Fitzpatrick:2015qma,Kaviraj:2015cxa,Alday:2014tsa,Alday:2015ewa,Alday:2015eya,Dey:2016zbg,Alday:2015ota,Kaviraj:2015xsa,Li:2015itl,Li:2015rfa}.

It may be interesting to systematically compute corrections to the leading-order behaviour~\reef{eq:leadingLC}. There is a group-theoretical approach to this problem, first discussed in appendix A of Ref.~\cite{Fitzpatrick:2015qma} (see also~\cite{Braun:2003rp}). We will briefly review their argument here. The idea is to restrict $SO(d,2)$ --- the conformal group in Minkowski signature --- to $SO(2,2)$, the group of conformal transformations acting on the $(z,\zb)$ plane. On the level of its Lie algebra, the latter splits into two copies of $\mfr{sl}(2)$, spanned by three chiral generators $L_0,L_{\pm 1}$ and three anti-chiral generators $\bar{L}_0,\bar{L}_{\pm 1}$. Any $SO(2,2)$ primary state is therefore labeled by two numbers $h,\bar{h}$, the eigenvalues of $L_0$ resp.\@ $\bar{L}_0$; such a state lifts to a local operator with scaling dimension $h + \bar{h}$ and spin $|h - \bar{h}|$.

Under this restriction, any $d$-dimensional conformal multiplet breaks up into infinitely many ``lightcone primaries''. As with the dimensional reduction discussed in this paper, this implies that any $d$-dimensional conformal block can be decomposed into 2$d$ blocks. Concretely, we have:
\beq
\label{eq:lcE}
G_{\DD}^{(\l)}(z,\zb;d) = \sum_{h,\bar{h}} P_{h,\bar{h}}(\DD,\l;d)\, G_{h+\bar{h}}^{(|h-\bar{h}|)}(z,\zb;2)
\eeq
for some coefficients $P_{h,\bar{h}}(\DD,\l;d)$ fixed by conformal symmetry. Every term in the RHS corresponds to a different lightcone primary with quantum numbers $(h,\bar{h})$. 
In the limit $z \to 0$, the sum~\reef{eq:lcE} is dominated by a single block with $h = (\DD - \l)/2$ and $\bar{h} = (\DD+\l)/2$, all other terms being suppressed by powers of $z$.

In order to compute corrections to~\reef{eq:leadingLC} it is therefore sufficient to determine the coefficients $P_{h,\bar{h}}(\DD,\l;d)$. This has not been done so far, to our knowledge. We remark however that the expression~\reef{eq:master} for the coefficients $\mca{A}_{n,j}(\DD,\l)$ is sufficient to determine the $P_{h,\bar{h}}(\DD,\l;d)$ for all integer $d$. For $d=3$ this is obvious: after relabeling $h,\bar{h}$ in the RHS of~\reef{eq:lcE} in terms of scaling dimensions and spins, the coefficients $P_{h,\bar{h}}$ are identical to the coefficients $\mca{A}_{n,j}$ with $d \to 3$.
The generalization to $d>3$ is straightforward: in order to determine the coefficients $P_{h,\bar{h}}(\DD,\l;d>3)$ one has to ``dimensionally reduce'' $d-2$ times.

\section{Discussion}\label{sec:disc}

This note has presented a new method to compute conformal blocks in $d$-dimensional CFTs, by relating them to conformal blocks of CFTs in $d-1$ dimensions. In particular, Eqs.~\reef{eq:genRep} and~\reef{eq:master} together form an explicit formula for blocks in odd $d$: for $d=3$ (resp.\@ $d=5$) our method leads to an expression in terms of 2$d$ (resp.\@ 4$d$) blocks shown in Eq.~\reef{eq:even}, which in turn are given by ${}_2F_1$ hypergeometric functions. Moreover, the expansion in lower-dimensional blocks converges exponentially fast, which may prove to be useful for numerical applications.

Currently only two closed-form expressions are known for conformal blocks in odd $d$: the $z$-series expansion~\reef{eq:serRep} and a formula that uses Mellin-Barnes integrals~\cite{mack1,mack2,Fitzpatrick:2011hu,DO3}. The latter involves so-called Mack polynomials that don't admit very compact expressions. The coefficients $\mca{A}_{n,j}$ from Eq.~\reef{eq:master} may therefore be easier to deal with in practice. In particular, they may be useful for the analytic bootstrap~\cite{Isachenkov:2016gim,Fortin:2016lmf} in three dimensions, since the two-dimensional crossing kernel is known in closed form~\cite{ana}.

There are a few obvious ways to extend the results presented in this note. First, it is possible write down a similar expansion for conformal blocks with non-zero external dimensions. The resulting expressions are somewhat more complicated, as the selection rule described below~\reef{eq:rhoSeries} does not apply. Second, it is possible to dimensionally reduce more complicated representations of the Lorentz group. A starting point for this would be the ``seed'' conformal blocks in three and four dimensions~\cite{Iliesiu:2015akf,Echeverri:2016dun}. An even further generalization consists of dimensionally reducing superconformal multiplets and the resulting superconformal blocks. We leave all of these issues for future work.

\subsection*{Acknowledgements}

We thank Carlo Meneghelli, Leonardo Rastelli and Martin Ro\v{c}ek for useful discussions.

\appendix

\section{Free field theory}\label{sec:free}

A consistency check of the results obtained in this paper is furnished by free scalar field CFT in $d$ dimensions. We recall that the four-point function of the free scalar $\phi$ is
\beq
\expec{\phi(x_1)\phi(x_2)\phi(x_3)\phi(x_4)} = \frac{\mca{G}_{\text{free}}(u,v)}{|x_{12}|^{d-2}|x_{34}|^{d-2}}\,, \qquad \mca{G}_\text{free}(u,v;d) = 1 + u^\nu + (u/v)^\nu\,.
\eeq
The above four-point function has a well-known CB decomposition, namely
\beq
\label{eq:ddec}
\mca{G}_\text{free}(u,v;d) = 1 + 2 \sum_{\l \text{ even}} f_\l \,  G_{\l + d-2 }^{(\l)}(u,v;d)\,, \qquad
f_{2p} = \frac{1}{c_{2p}^{(d)}} \frac{(\nu)_p (2\nu)_{2p}}{4^p\, (2p)! \,  (\nu-1/2+p)_p}\,.
\eeq

At the same time, we can decompose $\mca{G}_\text{free}(u,v)$ in terms of $d-1$ dimensional conformal blocks. From the $d-1$ dimensional point of view, the four-point function~\reef{eq:ddec} belongs to a free field theory with a non-local action, known as a generalized free field~\cite{ElShowk:2011ag}. The CB decomposition has the following form:
\beq
\label{eq:reddec}
\mca{G}_\text{free}(u,v;d) = 1 +  2 \sum_{\l \text{ even}} \sum_{n=0}^\infty g_{\l,n} \,  G_{\l + d-2 + 2n}^{(\l)}(u,v;d-1)\,.
\eeq
The coefficients $g_{\l,n}$ appearing here are given by~\cite{Fitzpatrick:2011dm}
\beq
g_{\l,n} =   \frac{c_\l^{(d)}}{c_{\l}^{(d-1)}} \, f_\l \, Z_\l^\l \times \begin{cases} 1 & n = 0\\
  2\la_n(\l) \quad & n \geq 1
\end{cases}\;,
\eeq
where we introduce the notation
\beq
\la_n(\l) \ldef  \frac{(1/2)_n}{16^n \, n!} \frac{(\l+2\nu-1)_{2n}(\l+\nu)_n}{(\l+\nu-1/2)_{2n}(\l+\nu+1/2)_n}\,.
\eeq

We want to verify that the CB decompositions~\reef{eq:ddec} and~\reef{eq:reddec} are consistent with the dimensional reduction formula~\reef{eq:master}. Notice that in~\reef{eq:ddec} only operators with twist $\tau = 0$ appear. In the zero-twist limit, the coefficients $\mca{A}_{n,j}$ simplify: 
\beq
\label{eq:Aatbound}
\mca{A}_{n,j}(\DD,\l) \; \limu{\tau \to 0} \; Z_\l^j\, \frac{c_\l^{(d)}}{c_j^{(d-1)}}   \times 
\begin{cases} \la_n(\l) \quad & j = \l \\
  \dd_{n,0} \quad & j < \l
  \end{cases}\,.
\eeq
Hence applying~\reef{eq:master} to Eq.~\reef{eq:ddec} gives the following CB decomposition in $d-1$ dimensions:
\beq
\mca{G}_\text{free}(u,v;d) = 1 +  2 \sum_{\l \text{ even}} \sum_{n=0}^\infty h_{\l,n} \,  G_{d-2 + \l + 2n}^{(\l)}(u,v;d-1)
\eeq
with coefficients 
\beq
h_{\l,0} =   \frac{c_\l^{(d)}}{c_\l^{(d-1)}}  \, Z_\l^\l  \, f_\l \,,
\qquad
h_{\l,n \geq 1} =  \frac{c_\l^{(d)}}{c_\l^{(d-1)}}  \,  Z_\l^\l \, f_\l \, \la_n(\l) +  \frac{c_{\l+2n}^{(d)}}{c_\l^{(d-1)}}\,  Z_{\l+2n}^\l  \, f_{\l+ 2n}\,.
\eeq
Consistency with~\reef{eq:reddec} requires that $g_{\l,n} = h_{\l,n}$ for all $\l,n$. For $n=0$ this is obvious, and for $n \geq 1$ this follows from the identity
\beq
c_{\l+2n}^{(d)} \, Z_{\l+2n}^\l \, f_{\l+2n} = c_\l^{(d)} \, Z_\l^\l \, f_{\l} \, \la_n(\l)\,.
\eeq

Similar consistency checks could be performed for more complicated four-point functions in free field or generalized free field CFTs.
{
\bibliographystyle{utphys}
\bibliography{biblio}
}

\end{document}